\documentclass[%
 reprint, 
 amsmath,
 amssymb,
 aps, 
 ]{revtex4-1}
\usepackage{hyperref}
\usepackage{xcolor}

\newcommand{\mbf}{\boldsymbol}

\usepackage{dsfont}

\usepackage{diagbox}

\usepackage{supertabular}

\newcommand{\ra}{\rangle}
\newcommand{\la}{\langle}

\newcommand{\hxi}{x_i}
\newcommand{\hxj}{x_j}
\newcommand{\hp}{p}

\newcommand{\hpi}{p_i}
\newcommand{\hpj}{p_j}
\newcommand{\hlk}{L_k}
\newcommand{\ijk}{\varepsilon_{ijk}}
\newcommand{\ijkL}{\ijk\hlk}

\newcommand{\magp}{|\mbf{\hp}|}

\newcommand{\commij}{[\hxi,\hpj]}
\newcommand{\commxx}{[\hxi,\hxj]}
\newcommand{\commpp}{[\hpi,\hpj]}
\newcommand{\ih}{i\hbar}
\newcommand{\dij}{\delta_{ij}}
\newcommand{\ie}{i.e., }
\newcommand{\eg}{e.g., }
\newcommand{\cf}{cf.\ }
\newcommand{\be}{\begin{equation}}
\newcommand{\ee}{\end{equation}}
\newcommand{\oo}[1]{\mathcal{O}({#1})}

\newcommand{\bs}[1]{\boldsymbol{#1}}
\newcommand{\bp}{\mbf{p}}

\newcommand{\brho}{\mbf{\rho}}
\newcommand{\bb}{\bs{\beta}}
\newcommand{\pbp}{\bp\cdot\bb\cdot\bp}
\newcommand{\pb}{\bp\cdot\bb}

\usepackage{soul}

\begin{document}

\preprint{APS/123-QED}

\title{Constraining GUP Models Using Limits on SME Coefficients}

\author{Andr\'e Herkenhoff Gomes}
 \email{andre.gomes@ufop.edu.br}
\affiliation{
 Departamento de F\'isica, Universidade Federal de Ouro Preto, Ouro Preto, MG, Brazil
 }

\begin{abstract}

Generalized uncertainty principles (GUP) and, independently, Lorentz symmetry violations are two common features in many candidate theories of quantum gravity. Despite that, the overlap between both has received limited attention so far. In this brief paper, we carry out further investigations on this topic. At the nonrelativistic level and in the realm of commutative spacetime coordinates, a large class of both isotropic and anisotropic GUP models is shown to produce signals experimentally indistinguishable from those predicted by the Standard Model Extension (SME), the common framework for studying Lorentz-violating phenomena beyond the Standard Model. This identification is used to constrain GUP models using current limits on SME coefficients. In particular, bounds on isotropic GUP models are improved by a factor of $10^{7}$ compared to current spectroscopic bounds and anisotropic models are constrained for the first time.
\end{abstract}


\maketitle

\textit{Introduction} --- Many tentative approaches to formulate a consistent quantum theory of gravity suggest a fundamental length scale in nature; among these are string theories, loop quantum gravity, and noncommutative geometry \cite{hagar,hossen2013}. Model-independent arguments indicate this fundamental scale may be revealed as departures from the Heisenberg uncertainty principle but cannot specify them completely \cite{maggiore,micro-bh}. Phenomenologically, implementation at the purely kinematic level of such departures can be done by modifying, or deforming, as commonly said, the canonical commutation relation $\commij=\ih\dij$ \cite{maggiore-algebra,kmm95}. This approach to generalized uncertainty principles (GUP)
began with investigations on the possible existence of a finite resolution for position measurements \cite{maggiore-algebra,kmm95} and has since grown far beyond, encompassing now, to name some, investigations on naturally occurring momentum cut-off at high energies \cite{kempf-mangano97,pedram12-plb2,pedram12-plb3}, existence of a classical regime at the Planck scale \cite{scardigli2010,petruzziello2020,maggiore2021}, and curved momentum space \cite{curved-momentum}.

It may be argued the so far most studied deformation of the canonical commutator is that proposed by Kempf \cite{kempf97},
\be
\commij = \ih [(1 + \beta p^2)\dij + 2\beta\hpi\hpj],
\ee
which predicts minimal uncertainty of $\hbar\sqrt{5\beta}$ for the measurement of any spatial Cartesian coordinate and features $\commxx = 0 +\oo{\beta^2}$. Analysis of the Lamb shift on the hydrogen $s$ level sets the upper bound for the parameter $\beta$ at around $10^{37}\ell^2_P/\hbar^2 \sim 0.1$ GeV\textsuperscript{$-2$} (in natural units) \cite{stetsko,Note1}%
\!\!\makeatletter{\renewcommand*{\@makefnmark}{}
\footnotetext{
There are stronger bounds reported in the literature (\cf Table II in \cite{igup}) that depend on extra assumptions regarding whether or how $\beta$ scales with the particle number of composite bodies, but these are not completely well-understood at the moment and may affect such bounds by several orders of magnitude  \cite{composite}.
}\makeatother}
and places this sort of GUP prediction in the regime of nuclear physics.

At the nonrelativistic level, to the best of our knowledge, investigations so far concentrate on GUP models in isotropic scenarios (iGUP) --- see \cite{igup} for a recent review. Even in the context of, \eg $\kappa$-deformed Poincaré algebra, the notion of an isotropic deformation parameter is retained \cite{kappa}. Nevertheless, it is unclear whether spacetime necessarily exhibits such symmetric behavior at energy scales relevant to quantum gravitational phenomena. Assuming it may not, we report efforts toward extending the study of GUP to anisotropic scenarios. The possibilities are many, but for an illustration, consider some anisotropic versions of Kempf's model:
\be
\commij \stackrel{\text{?}}{=} \ih \left\{
\begin{array}{l}
 [1+(\bb\cdot\mbf{\hp})^2 ]\dij + \cdots, \medskip\\
 (1 + \pbp)\dij + \cdots, \medskip\\
 \qquad \dots
\end{array}
\right.
\ee
Ignoring differing dimensionalities, we see the isotropic deformation $\beta\hp^2$ is generalized to more complex structures coupling the momentum to a vector, as in $(\bb\cdot\mbf{\hp})^2$, or to a second-rank tensor, as in $\pbp$, among others, but the basic idea is that the scalar nature of the deformation parameter $\beta$ is generalized to that of a background tensor $\bb$ effectively introducing preferred directions in space. It turns out, such spatial anisotropy is a hallmark of so-called Lorentz symmetry violations, signals of which have been intensively studied in the last decades within the Standard Model Extension (SME) framework \cite{sme1,sme2,sme3} and experimentally searched \cite{datatable} as candidate signatures of quantum gravity \cite{tasson,bluhm}. Despite both GUP and Lorentz-violating (LV) phenomena being possible encounters on the road to a consistent quantum theory of gravity, their overlap has been little explored so far. A noteworthy exception is found on \cite{gup-sme}, where the authors explore the overlap between the gravitational sector of the SME and iGUP physics. In this paper, we focus instead on the fermion sector of the SME and GUP models in anisotropic space (aGUP).

Physics from modified canonical commutators in anisotropic scenarios has nevertheless been investigated in at least two other approaches \cite{petrov-book}, so it is worth briefly placing the GUP approach into a wider context.

One is noncommutative quantum field theory (NCQFT) \cite{ncqft}. In a simpler form, it is based on quantum fields depending on noncommuting position coordinates satisfying $[x^\mu,x^\nu]=i\theta^{\mu\nu}$, with $\theta^{\mu\nu}$ an antisymmetric constant tensor ultimately breaking Lorentz symmetry \cite{ncqft-sme}. The other approach hinges on deforming the commutator of field momenta $\pi_i$ in the Hamiltonian formalism by setting, \eg $[\pi_i(\mbf{x}),\pi_j(\mbf{y})] = \theta_{ij}\delta^3(\mbf{x}-\mbf{y})$ instead of zero \cite{qft-deformed1,qft-deformed2,qft-deformed3}. Both approaches provide a mechanism to produce Lorentz-violating Lagrangians formally equivalent to a subset of those of the SME. Quite notably, for the case of gauge fields, the former approach leads to CPT invariant physics \cite{ncqft-sme} while the latter to CPT-violating one \cite{qft-deformed1,qft-deformed2}.

The GUP approach, when extended to anisotropic space as discussed in this paper, is found to offer another mechanism for Lorentz-violating physics that matches with a subset of that contained in the nonrelativistic limit of the SME and is particularly suited when considering nonrelativistic scenarios such as spectroscopic experiments. A question to be answered by future investigations is whether it is a truly independent approach. For one reason, the nonrelativistic quantum mechanical limit of NCQFT \cite{ncqft-qm1,ncqft-qm2,ncqft-qm3} seemingly does not relate to GUP in any trivial way, since the latter is based on momentum-dependent deformations of $\commij$ and needs not feature noncommutativity of position coordinates. For another, comparison to quantization based on deformed commutators for fields requires developing a quantum field model based on such aGUP (possibly along the lines of, \eg \cite{gup-qft,bosso-luciano}), something beyond our current scope.

Next, we establish the context of this paper regarding GUP and the SME framework. Then, after a brief overview of key aspects of iGUP models, we show that, at the level of effective physics, predictions from a large class of iGUP Hamiltonians are indistinguishable from those of the nonrelativistic SME Hamiltonian for fermions. This result highlights that iGUP does not necessarily originate from Lorentz-invariant physics and motivates consideration of more general aGUP models. We tackle this case in the paper's second half, comparing its predictions to those of the SME and extracting realistic bounds on both isotropic and anisotropic GUP parameters.

\textit{The SME and GUP} --- Even though the SME is based on conventional commutators, its Lagrangian is constructed to contain all possible terms (Lorentz-invariant or not) correcting the Standard Model (SM) plus General Relativity (GR) Lagrangian at the classical level. Even though the discussed approach to GUP models modifies quantum mechanics at its heart, any realistic prediction must come at the level of effective physics as small corrections to the SM plus GR as well. The expectation, therefore, is that any GUP-based prediction is contained in the SME framework. Full exploration of this claim requires a quantum field theoretical approach entailing predictions of deformed canonical commutators at the nonrelativistic limit and is beyond our scope. Instead, we focus on showing that at least a class of nonrelativistic GUP models, specified next, predicts Hamiltonians equivalent to those of the nonrelativistic SME fermion sector.

In flat spacetime, SME coefficients controlling interactions beyond the Standard Model are usually taken as constants in both space and time, behaving as fixed background tensors. This assumption enforces invariance under spacetime translations at the fundamental level and thus energy-momentum conservation; hence, we consider only nonrelativistic GUP models featuring the same, which are those with commuting position operators ($\commxx=0$).  This restriction is taken here for simplicity, but should be relaxed in future works, which may involve comparing GUP models with non commuting coordinates to the curved spacetime formulation of the SME \cite{sme3,kost-zonghao1,kost-zonghao2}.

At last, analysis of the SME fermion sector reveals that, at the level of effective physics, isotropic corrections to the Hamiltonian are necessarily spin-dependent when accompanying odd powers in the momentum --- \cf Eqs.\  (91) and (92), and Table III of \cite{kost-mewes-fermion}. In contrast, a long-standing challenge has been formulating GUP models for spin or within a sensible relativistic regime where spin manifests naturally; thus, we are led to consider only commutator deformations based on even powers in the momentum --- but we remark that this simple comparison suggests any realistic isotropic GUP model with a Hamiltonian containing odd powers in the momentum should be spin-dependent. To keep our discussion as general as possible, this particular restriction is usually enforced only when writing down Hamiltonians.


\textit{Isotropic GUP models} --- Demanding spatial isotropy restricts the most general deformation of the canonical commutation relation to be
\be
\label{ccr}
\commij = \ih[ f(p) \dij + g(p) \hpi\hpj]
\ee
as long as the momenta are kept conventional, $\commpp=0$, with $f$ and $g$ real functions of $p\equiv\magp$ and deformation parameters (\eg $\beta$). Depending on these, position measurements cannot be made arbitrarily precise and, as a result, position eigenstates no longer form a basis for physical states \cite{kmm95,pedram12-prd,pasquale-xbasis}; formally $\hxi$ is no longer self-adjoint but merely symmetric \cite{kempf2000}. This possibility is excluded for momentum measurements and the deformed algebra finds a Hilbert space representation on momentum wave functions, ${\psi}(\bp) = \la\bp|\psi\ra$, with momentum represented by the conventional multiplicative operator and position represented by
\begin{align}
\hxi {\psi}(\bp) & = \ih(f \partial_{p_i} + g\hpi\bp\cdot\mbf{\partial}_{\bp} + \gamma\hpi) {\psi}(\bp),
\end{align}
with $\gamma(p)$ unobservable and adjustable to ensure $\hxi$ is a symmetric operator. Additionally, position operators do not necessarily commute,
\be
\label{ccrxx}
\commxx = \ih [g ( f - p \partial_p f ) -  p^{-1} f \partial_p f] \ijkL,
\ee
where $\hlk \equiv f^{-1}\ijk\hxi\hpj$ holds whenever spin degrees of freedom are neglected and is identified as the generator of three-dimensional rotations \cite{maggiore2021}. One particular consequence of $\commxx\neq0$ is the loss of translation invariance at the fundamental level, even though rotational symmetry is retained. Since we are not interested in this feature, we set a commutativity condition relating $f$ and $g$,
\be
gp = \frac{f \partial_p f}{f - p \partial_p f}
\quad
\longleftrightarrow
\quad
\commxx = 0.
\ee
For such commutative GUP models, there is a generator of translations, $\mbf{T}=\bp/f(p)$, where $[\hxi,T_j]=\ih\dij$ holds. Even though it can no longer be identified as the momentum operator, it still provides a wave function representation \cite{kempf-mangano97}, $\tilde{\psi}(\brho) = \la \brho|\psi\ra$ where $T_i|\brho\ra=\rho_i|\brho\ra$, such that
\be
\hxi \tilde{\psi}(\brho) = \ih\partial_{\rho_i} \tilde{\psi}(\brho),
\ee
\be
\hpi \tilde{\psi}(\brho) = \hpi(\brho) \tilde{\psi}(\brho)
\quad
\longleftrightarrow
\quad
\frac{p_i}{f(p)} = \rho_i,
\ee
where at small $\rho_i$ one has $p_i(\brho)\approx\rho_i$. From here on, we will discuss iGUP within this representation. The main advantage is that of retaining the conventional, simpler derivative structure of the position operator while encoding the physics of deformed commutators in the relation between $\bp$ and $\brho$ --- at least for the one-dimensional case, it is known the model features nonvanishing minimum position uncertainty if and only if $p_i(\brho)$ is defined only for $\rho_i$ belonging to a restricted interval \cite{mangano2016}. Besides providing the foundations for what comes next, this brief overview of the algebraic formulation of iGUP will also work as a reference for the analogous development of the anisotropic GUP we do later.

Since effective physics from GUP is expected to be suppressed by powers in the inverse of the Planck mass, a perturbative approach is favored when considering possible  experimental signals. Expressing $p_i(\brho)$ as a power series in $\rho$ and restricting to $p_i(\rho^2) = \rho_i(1 + \sum \alpha_{n}\rho^{n})$ ($n$ even $\ge0$), the canonical commutation relation reads
\be
\commij = \ih \left( \overline{\alpha}_0 + \frac{\alpha_2}{\overline{\alpha}_0^2} p^2 \right) \dij + \frac{2\alpha_2}{\overline{\alpha}_0^2} \hpi \hpj + \oo{p^4},
\ee
with $\overline{\alpha}_0=1+\alpha_0$, where we explicitly show terms only up to $\oo{p^2}$ for clarity. Since $\alpha_0$ is dimensionless, it must depend on deformation parameters such that $\alpha_0\to0$ if these parameters vanish. 
To illustrate an interesting consequence of $\alpha_0\neq0$, consider there is a single parameter $\beta$ of dimension $[\beta]=[p]^{-2}$ so that $\alpha_0=\text{const.}\times\beta$. Dimensional analysis reveals the constant may depend on the particle's mass, allowing deformations of the canonical commutator to be dependent on the particle species; one possibility is through a species-dependent mass rescaling of $\hbar$ (\eg see commutator (13) on \cite{maggiore2021}). In the spirit of commutator deformations as universal phenomena, $\alpha_0=0$ is often set --- and we remark $\alpha_2$ above is then identified as $\beta$ in the context of Kempf's model mentioned before. Here $\alpha_0\neq0$ is chosen instead for greater generality.

Following this perturbative approach, the free fermion Hamiltonian splits into a conventional kinetic part $\rho^2/2m_\psi$ and a perturbation
\be
\delta H_\text{iGUP} = \frac{ 2\alpha_0+\alpha_0^2 }{ 2m_\psi } \rho^2 + \sum_{ n=4,6,\dots }^\infty \sum_{k=0,2,\dots}^{n-2} \frac{ \overline{\alpha}_{n-k-2} \overline{\alpha}_{k} }{ 2m_\psi } \rho^{n}
\ee
with $\overline{\alpha}_n=\alpha_n$ for $n>0$. As we see next, each $\alpha_n$ corresponds to a specific isotropic nonrelativistic SME coefficient and can be bounded accordingly.


\textit{Comparison to the isotropic nonrelativistic SME Hamiltonian} --- The free-fermion sector of the SME has been studied in great details in \cite{kost-mewes-fermion}. There, the full Lorentz-violating perturbation Hamiltonian is structured in Eq.\ (80) in Cartesian basis and in (88) in spherical basis. Neglecting spin-dependent perturbations, the nonrelativistic isotropic limit is conveniently given in spherical basis by (113), here written as
\be
\delta H_\text{iLV} = \sum_{n=0,2,\dots}^\infty  ( \mathring{a}_{n}^\text{NR} - \mathring{c}_{n}^\text{NR} )\rho^{n},
\ee
where $\mathring{a}^\text{NR}_{n}$ and $\mathring{c}^\text{NR}_{n}$ are isotropic Lorentz-violating SME coefficients of mass dimension $1-n$, the first associated to CPT-violating physics as well. Originally, this expression involves $\bp$ instead of $\brho$, but since the SME is based on conventional quantum mechanics, we can identify both as the same quantity here. 

Isotropic GUP models, as formulated by deformations of the canonical commutation relation, respect charge, parity and time reversal symmetries and thus relate only to SME coefficients $\mathring{c}_{n}^\text{NR}$ above. Among these, the leading effect of $\mathring{c}_{0}^\text{NR}$ is to equally rescale all energy eigenvalues in the preferred inertial frame where this isotropic SME limit holds while inducing anisotropic signals in other frames \footnote{To leading-order in SME coefficients, see, for instance, Hamiltonian (2) of \cite{brett-c00} in the context of hydrogen electronic transitions.}. Since iGUP models are constructed under the premise of spatial isotropy, there is no GUP parameter analogous to $\mathring{c}_{0}^\text{NR}$ or, equivalently, one could say it is already absorbed in the definition of the energy eigenvalues. Finally, the overlap between $\delta H_\text{iGUP}$ and $\delta H_\text{iLV}$ corresponds to identifications
\be
2\alpha_0+\alpha_0^2 = - 2m_\psi \mathring{c}_{2}^\text{NR},
\ee
\be
\sum_{k=0,2,\dots}^{n-2} \overline{\alpha}_{n-2-k} \overline{\alpha}_k = - 2m_\psi \mathring{c}_{n}^\text{NR}
\quad
(n\text{ even }\ge4),
\ee
confirming that predictions from commutative iGUP models based on deformed commutators with $f=f(p^2)$ are contained in the SME framework.

A wide range of SME coefficients have been constrained in the past decades by state-of-the-art high precision experiments \cite{datatable} and this offers a great opportunity for deriving new constraints on GUP models. The available experimental bound for $\mathring{c}_{n}^\text{NR}$, in particular, is in the electron sector and comes from an observed $1.8$ $\sigma$ difference in the theoretical and experimental values of the Positronium 1S-2S frequency \cite{kost-arnaldo}:  
\be
\mathring{c}_{2}^\text{NR} + \tfrac{67}{12}(\alpha m_\text{r})^2\mathring{c}_{4}^\text{NR} \simeq (4.5 \pm 2.4) \times 10^{-6} \text{ GeV}^{-1},
\ee
where $\alpha \approx 1/137$ is the fine-structure constant and $m_r = \frac{1}{2}\times0.511$ MeV is the Positronium reduced mass, suggesting experimental reach of about $10^{-5}\text{ GeV}^{-1}$ for $\mathring{c}_{2}^\text{NR}$ and $10^{5}\text{ GeV}^{-3}$ for $\mathring{c}_{4}^\text{NR}$. Taken independently, it sets the following constraints on iGUP parameters:
\be
|\alpha_0| \lesssim 10^{-8}, 
\quad
|\alpha_2| \lesssim 10^{2}\text{ GeV}^{-2} \sim 10^{40} \ell_P^2/\hbar^2.
\ee
The constraint on $\alpha_0$ is, as far as we know, new to the literature. Although the bound on $\alpha_2$ is weaker than current spectroscopic bounds ($\sim0.1 \text{ GeV}^{-2}$ \cite{stetsko}), we will see in the following that taking into account anisotropic GUP and the available limits on other SME coefficients allows much stronger bounds.


\textit{Anisotropic GUP models} --- If Lorentz symmetry is not an exact symmetry of nature, the expression for $\delta H_\text{iGUP}$ should be understood as holding only in preferred inertial reference frames while invalid in others as sequences of boosts generally induce a relative rotation among frames. Thus, if GUP has its origin in high-energy Lorentz-violating phenomena, its isotropic formulation predicts the subset of rotation-invariant effects only. A wealthier set of predictions is then achieved considering anisotropic formulations of GUP.

The commutativity of momentum operators is an automatic feature in the isotropic scenario, but this is not so for the anisotropic case. Nonetheless, for a less drastic departure from isotropy we assume $[\hpi,\hpj]\equiv 0$. By all means, giving up spatial isotropy still allows the canonical commutation relation to have virtually any tensor structure depending on what sort of anisotropies are considered. What we propose is the following anisotropic deformation:
\be
\commij = \ih[ f(\bp)\dij + g_i(\bp)\hpj],
\ee
with $f$ and $g_i$ real. Notice the isotropic case is immediately recovered setting $f(\bp)\to f(p)$ and $g_i(\bp)\to g(p)\hpi$. Despite not exhausting all the possibilities for anisotropic deformations, this proposal is found to allow for a very direct generalization of the commutativity condition \footnote{Actually, we find that at least for a slightly more general deformation $\commij=\ih[f(\bp)\dij+g_i(\bp)h_j(\bp)]$, there is no condition enforcing commutativity of position operators unless we set $h_j=\hpj$ and recover our proposal. Whether the same conclusion can be extended to the general case $\commij=\ih F_{ij}(\bp)$ is unclear to us.}; namely,
\be
g_i = \frac{f \partial_{p_i} f}{f - \bp\cdot\mbf{\partial}_{\bp} f}
\quad
\longleftrightarrow
\quad
\commxx = 0,
\ee
which can be verified by straightforward calculation noticing the deformed algebra can be realized on momentum wave function representation setting
\begin{align}
\hxi {\psi}(\bp) & = \ih(f \partial_{p_i} + g_i\bp\cdot\mbf{\partial}_{\bp} + \gamma_i) {\psi}(\bp),
\end{align}
with $\gamma_i(\bp)$ unobservable and freely adjustable, as in the isotropic case, to ensure $\hxi$ is symmetric. From here and on, we consider this commutative case only.

The function $f(\bp)$ is dimensionless, hence depends on momentum only through the combination $\beta_{i_1 i_2\dots i_n}p_{i_1} p_{i_2} \cdots p_{i_n}$, with $\beta_{i_1 i_2\dots i_n}$ a generic background field of dimension $[p]^{-n}$. Considering the behavior of $i\mbf{\partial}_{\bp}$ and $\bp$ under the discrete transformations of parity and time reversal, we infer that coupling to background tensors of even rank preserve both symmetries, whereas coupling to those of odd rank violate them; further inspection also reveals $g_i(\bp)$ is an odd function under parity or time reversal transformations. The theory is therefore allowed to violate discrete symmetries at the fundamental level and not only by external potentials (\eg particle decay), but since we are considering only deformations of the canonical commutation relation containing even powers in the momentum, all discussion that follows thus assumes such discrete symmetries are preserved.

Almost identically to the isotropic case, the anisotropically deformed algebra finds a representation on eigenvectors of the translation operator $T_i$ and expressions for $\hxi$ and $\hpi$ are structurally the same as before, with the exception of $p_i/f(\bp) = \rho_i$, because $f$ is allowed to be direction-dependent. Generally, such dependence prevents solving exactly for $p_i(\brho)$ and a perturbative solution may be needed, but for any particular model featuring a single anisotropy, an exact solution is always attainable in principle; at the end of the day, one is ready to write down the Hamiltonian.

Due to the complexity of explicitly writing down a general expression for $f(\bp)$, to the rest of this paper we concentrate on a specific model, though we remark that our methods apply to any other as long as $f(\bp)$ contains only even powers in the momentum. Hence, consider one possible anisotropic version of Kempf's model:
\be
\commij = \ih[(1+\pbp)\dij + g_i(\bp) p_j],
\ee
with commutativity of position operators imposing
\be
g_i=2(\pb)_i\frac{1+\pbp}{1-\pbp}.
\ee
The isotropic case is recovered from the 6 linearly independent (LI) components of $\bb$ by restricting to the only scalar component, $\beta_{ij} \to \beta\dij$. In this case, $g_i\approx 2(\pb)_i$ reduces to $g p_i\approx2\beta p_i$ as expected \cite{kempf97}. 

Finding $p_i(\brho)$ is immediate after solving $p_i/f(\bp)=
\rho_i$ for $\pbp$ first, then picking the right solution by demanding $p_i \to \rho_i$ for $\bb\to0$. One finds
\be
\hpi(\brho) = \frac{ 2 \rho_i }{ 1 + \sqrt{ 1 - 4\brho \cdot \bb \cdot \brho } }.
\ee
An interesting feature of this model is the restriction $\brho\cdot\bb\cdot\brho \le 1/4$ on translation eigenvalues or, equivalently, $\bp\cdot \bb \cdot \bp \le 1$ on momentum eigenvalues, effectively setting a direction-dependent upper limit to the momentum (as opposed to a single upper limit on its absolute value as in isotropic models \cite{pedram12-plb2,petruzziello2020}). Although this behavior motivates further investigation, it is beyond our present scope since we are mostly concerned with relating this model to LV-models. For this intent, we write the free fermion perturbation Hamiltonian to first order on $\bb$:
\be
\delta H_\text{aGUP} = \brho\cdot\bb\cdot\brho \frac{\rho^2}{m_\psi} = \sum_{\ell m} \frac{\beta_{4\ell m}}{m_\psi} Y_{\ell}^m (\hat{\brho}) \rho^4.
\ee
where $Y_\ell^m(\hat{\brho})$ denotes the spherical harmonic function of degree $\ell$ and order $m$, with the unit vector $\hat{\brho} = (\sin\theta \cos\phi, \sin\theta \sin\phi, \cos\theta)$ defining the usual spherical polar angles; and we also define a spherical coefficient $\beta_{4\ell m}$ corresponding to the Cartesian $\beta_{ij}$ aiming further comparison to SME coefficients. The subindex $4$ is a reminder $\beta_{4\ell m}$ comes along with $\rho^4$, also $\ell=\{0,2\}$ to ensure $\beta_{4\ell m}$ has 6 LI components as $\beta_{ij}$ does.


\textit{Comparison to the nonrelativistic SME Hamiltonian} --- The nonrelativistic SME Hamiltonian is now considered in its general anisotropic version, Eq.\ (108) of \cite{kost-mewes-fermion}. Its spin-independent part (109) is given in terms of spherical SME coefficients due to their simpler rotation properties and in our notation reads
\be
\delta H_\text{LV} = \sum_{n \ell m} ( a^\text{NR}_{n\ell m} - c^\text{NR}_{n\ell m} )Y_{\ell}^m (\hat{\brho}) \rho^n,
\ee
where $a^\text{NR}_{n\ell m}$ and $c^\text{NR}_{n\ell m}$ are the relevant coefficients controlling Lorentz violation at this level. Note $\ell=m=0$ components correspond, up to a factor of $\sqrt{4\pi}$, to isotropic coefficients $\mathring{a}^\text{NR}_{n}$ and $\mathring{c}^\text{NR}_{n}$ discussed before, sharing corresponding behavior under CPT. The summation ranges are $n\ge0$, $\ell = n, n-2, n-4, {\dots} \ge0$, and $|m|=\ell, \ell-1, {\dots} \ge 0$. The number of independent components for both coefficients is $\frac{1}{2}(n+1)(n+2)$ and the mass dimension is $1-n$.

To find the correspondence of SME coefficients and aGUP parameters, we begin retaining only $c_{n\ell m}^\text{NR}$ above to ensure invariance under CPT transformation. Next, remember the anisotropy $\beta_{ij}$ has 6 LI components and comes along with $\rho^4$ in $\delta H_\text{aGUP}$. In contrast, $c_{2\ell m}^\text{NR}$ has 6 LI components, but in $\delta H_\text{LV}$ couples to $\rho^2$ instead. On the other hand, $c_{4\ell m}^\text{NR}$ couples to $\rho^4$ as desired, despite having 15 LI components among $\ell=4$ (9 LI), $\ell=2$ (5 LI), and $\ell=0$ (1 LI). This suggests $\beta_{ij}$ is completely contained in $c_{4\ell m}^\text{NR}$ restricted to $\ell=\{0,2\}$. The exact correspondence is
\be
\beta_{4\ell m} = m_\psi c_{4\ell m}^\text{NR}
\quad
\text{for}
\quad
\ell=\{0,2\},
\ee
revealing Hamiltonian predictions from our generalization of Kempf's model to anisotropic commutator deformations are contained in the SME framework as well.

Even though no direct bound on the SME coefficient $c_{4\ell m}^\text{NR}$ is currently available, we can still estimate upper limits for components of $\beta_{ij}$ after the above identification. To begin, we note the nonrelativistic coefficient $c_{n\ell m}^\text{NR}$ is a mix of relativistic coefficients $c^{(d)}_{n\ell m}$ with different mass dimensions ($4-d$) related by Eq.\ (111) of \cite{kost-mewes-fermion}. The same arguments as before indicate all components of $\beta_{4\ell m}$ are contained in such mixture restricted to coefficients with $n=4$ only, namely 
\be
\beta_{4\ell m} = \sum_{d=6,8,\dots}^\infty m_\psi^{d-6} c^{(d)}_{4\ell m}
\quad
\text{for}
\quad \ell=\{0,2\}.
\ee
In passing, because $d$ is higher than 4, we remark that only nonrenormalizable operators in Lorentz-violating quantum field theory produce nonrelativistic effects mimicking this particular deformation of the canonical commutator, be it isotropic or not. 

To proceed and estimate the experimental sensitivity to $\beta_{ij}$, we assume the least suppressed contribution to $\beta_{4\ell m}$ comes from the spherical coefficient $c^{(6)}_{4\ell m}$. At the present, high precision measurements of the 1S-2S hydrogen transition frequency offers the best constraint on components of the corresponding Cartesian coefficient $c^{(6)\mu\nu\rho\sigma}_\text{eff}$ defined on Eq.\ (27) of Ref.\ \cite{kost-mewes-fermion}. Since the laboratory frame where these measurements took place is non-inertial due to Earth's rotation, any fixed background field would have appeared time-dependent, making comparison of measurements on different time periods unattainable. To address this issue, bounds on spatial anisotropies are widely reported in the canonical Sun-centered frame \cite{sun-centered,datatable} with coordinates ($T,X,Y,Z$), which in the timescale of laboratory experiments can be taken as inertial. In this frame, searches for annual variations of the 1S-2S frequency generally places
\be
c_\text{eff}^{(6)TTTJ} + \sum_{K=1}^3 c_\text{eff}^{(6)TKKJ} \lesssim 10^{-4}\text{ GeV}^{-2}
\ee
where $J=\{X,Y,Z\}$ --- the exact bound varies for each $J$ \cite{kost-arnaldo}, but as we look for an estimate on $\beta_{ij}$, the weakest suffices. Following that, we identify $c^{(6)\mu\nu\rho\sigma}_\text{eff}$  with $\beta^{(\mu\nu}\eta^{\rho\sigma)}$, where $\beta^{ij}=\beta_{ij}$ are the only nonvanishing components, the spacetime metric is $\eta^{\mu\nu}=\text{diag}(+,-,-,-)$, and enclosing parentheses mean symmetrization on all indices. Under this assumption,
\be
\beta^{TJ} \lesssim 10^{-4} \text{ GeV}^{-2} \sim 10^{34} \ell^2_P/\hbar^2
\ee
in the Sun-centered frame. The derivation of this constraint has a far-reaching significance: even though we start from a nonrelativistic model for anisotropic GUP, we are still able to derive bounds on parameters that would otherwise appear only in a relativistic formulation. This is possible because the effective physics of GUP with $\beta_{ij}$ is fully contained in the SME framework, and boosting to the Sun-centered frame mix spatial and temporal indices.

The bound on $\beta^{TJ}$ places effects of this particular parameter in the realm of nuclear physics and may look at first underwhelming as current particle physics experiments probe matter in finer details, \eg at energy scales of $\sim 10\text{ TeV}$ at the Large Hadron Collider. Nevertheless, it should be noted that bounds reported in the Sun-centered frame are suppressed due to Earth's orbital speed of about $10^{-4}$ in natural units. It turns out the bound on the above SME coefficient combination suggests experimental reach of $10^{-8}\text{ GeV}^{-2}$ to the isotropic combination $\mathring{c}^{(6)}_2=c_\text{eff}^{(6)0011}+c_\text{eff}^{(6)0022}+c_\text{eff}^{(6)0033}$ in the laboratory frame (\cf Eq.\ (98) of \cite{kost-mewes-fermion}, Eq.\ (63) and Table XVIII of \cite{kost-arnaldo-clock}). As a consequence, we derive the constraint
\be
\beta_{11} + \beta_{22} + \beta_{33} \lesssim 10^{-8} \text{ GeV}^{-2} \sim 10^{30} \ell^2_P/\hbar^2,
\ee
placing possible experimental signals around the 10 TeV scale in the laboratory frame. Assuming there is a preferred direction in space, we remark that this coefficient combination is isotropic in the laboratory frame only at a uniquely specified position in space and time while being anisotropic at any other as Earth rotates and moves around the Sun; thus, the constraint reported in the Sun-centered frame is physically more meaningful. If, on the other hand, space is isotropic and $\beta_{ij} = \beta\delta_{ij}$, \ie from a different perspective, assuming Lorentz symmetry invariance, the constraint
\be
\beta \lesssim 10^{-8} \text{ GeV}^{-2} \sim 10^{30} \ell^2_P/\hbar^2
\ee
is frame-independent, and represents improvement by a factor of $10^{7}$ over current spectroscopic bounds on the parameter $\beta$ of isotropic GUP (\cf \cite{stetsko}).

Our results highlight a profitable approach for constraining GUP models, both Lorentz-violating and Lorentz invariant; namely, establishing a link between GUP parameters and SME coefficients. A similar idea was employed in \cite{gup-sme}, although with two key differences worth further discussion.

The first is the approach itself: instead of comparing Hamiltonians, the authors compare corrections to the Hawking temperature coming from GUP in isotropic space and those from Lorentz symmetry violation in the gravitational sector of the SME. The bound they derive for $\beta$ is of gravitational nature while ours is electromagnetic; thus both appear as complementary to each other. Bounding GUP by such comparisons to the SME is profitable indeed: the authors of \cite{gup-sme} report improvement by a factor of $ 10^9$ over previous gravitational bounds, setting $\beta \lesssim 10^{51}\ell^2_P/\hbar^2$.

The second key difference is that the authors of \cite{gup-sme} consider GUP in isotropic space only. One of their targets is deriving corrections to the Hawking temperature depending on the parameter $\beta$. For such, they start from the simplest GUP --- $\Delta x \Delta p \ge \tfrac{\hbar}{2}[1+\beta(\Delta p)^2]$ in our notation --- without resorting to any commutator representation for position and momentum, then follow heuristic arguments to relate GUP to the Schwarzschild radius and finally to the Hawking temperature. A remarkable feature of this approach to GUP and gravity is that it evades issues with violations of the equivalence principle (for more details, see \cite{scardigli-casadio,casadio-scardigli-2020}). The limitation to GUP in isotropic space may be traced back to the effectively one-dimensional approach (\ie along the radial direction only), but in the light of GUP as another possible source of Lorentz violation, as proposed in this paper, it could be interesting to extend their approach to GUP in anisotropic space.

\textit{Conclusions and Perspectives} --- Physics of anisotropic GUP models offers a fresh venue for investigation of possible quantum gravitational phenomena. Here we reported some steps on formulating such models in the case of $\commxx=\commpp=0$ with deformations based on even powers in the momentum. We investigated its overlap with Lorentz-violating models in flat spacetime in the SME framework, showing both predict equivalent effective physics. An intermediate byproduct was the insight that, to be consistent with the SME, the Hamiltonian for isotropic GUP models with commuting coordinates should be spin-dependent whenever containing odd powers in the momentum. From the overlap of SME and GUP, our main result was the derivation on new bounds on GUP parameters; in particular, improvement by a factor of $10^7$ on $\beta$ of isotropic GUP and stablishment of the first constraints on anisotropic parameters.

There are at least three interesting directions for future investigations. One is the extension of our results to GUP models with $\commxx\neq0$, where translation symmetry is broken and connection with the SME may require its formulation in curved spacetimes \cite{sme3,kost-zonghao1,kost-zonghao2}. The second is the formulation of GUP in the presence of spin degrees of freedom, which was recently revisited in the context of GUP with noncommutative coordinates \cite{maggiore2021}. The last one is the actual meaning of an anisotropic canonical commutator; in particular, as in the generalization $1+\beta p^2 \to 1+\pbp$ of Kempf's model, we may expect anisotropic commutators to generally allow for relative directions between momentum and anisotropy where $\commij\to0$ at sufficiently high momenta. Although the idea of a classical regime at the Planck scale has been investigated in the past years \cite{scardigli2010,petruzziello2020,maggiore2021}, linking it to preferred directions in space, and possibly time, has not been considered yet and may introduce a variety of interesting phenomena that might help to sort out models producing sensible physics.

\textit{Acknowledgments} --- The author wish to thank Alan Kosteleck\'y for very useful remarks during the conceptual stage of this work and also the anonymous referees for the interesting remarks and for pointing out very relevant references.

%
\bibliography{references}

\end{document}